\input{aipcheck}

\documentclass[
  ,final            
  ]
  {aipproc}

\layoutstyle{6x9}

\newcommand{\bea}{\begin{eqnarray}}
\newcommand{\eea}{\end{eqnarray}}
\newcommand{\be}{\begin{equation}}
\newcommand{\ee}{\end{equation}}

\newcommand{\lnum}{{\cal L}}

\newcommand{\til}{\widetilde}
\newcommand{\br}{{\tt Br}}

\newcommand{\fb}{\rm fb}

\begin{document}

~~~~~~~~~~~~~~~~~~~~~~~~~~~~~~~~~~~~~~~~~~~~~~~~~~~~~~~~~~~~~~~~~~~~~~~~~~~~~~~~~~~~~~~~~~~~~~~~~~~~~~~~~~~~~~~~~~~~~~~~~~~~~~~~~~~~~~~~~~~~~~~ {\small \tt UCRHEP-T475}\\[-10mm]

\title{New physics search at the LHC via $Z'$ resonance}



\author{Hye-Sung Lee}
{
address={Department of Physics and Astronomy, University of California, Riverside, CA 92521, USA}
}



\begin{abstract}
A new Abelian gauge symmetry $U(1)'$ is well motivated to extend the standard model or supersymmetric standard model.
Leptonic $Z'$ resonances are clean signals even at the hadron colliders.
We discuss how 4-lepton and 6-lepton $Z'$ resonances can help the search for Supersymmetry and Higgs, respectively, at the LHC.
\end{abstract}

\maketitle


\section{Introduction}
This talk\footnote{The talk was given at SUSY 2009 conference in Boston, MA.} is mostly based on a few recent papers \cite{Lee:2008cn,Barger:2009xg}.

An extra Abelian gauge symmetry $U(1)'$ at TeV scale is predicted in many new physics scenarios. (For a recent review, see Ref.~\cite{Langacker:2008yv} as well as Talk~\cite{Langacker:2009im}.)
The first implication of the $U(1)'$ for the LHC is a new gauge boson $Z'$ at TeV scale.
Although there are other discovery goals at the LHC (with arguably better motivations) such as Higgs and Supersymmetry (SUSY), $Z'$ is very likely to be the first discovery at the LHC if it exists.

$Z'$ can decay into a light lepton ($\ell = e$, $\mu$) pair, and it is very easy to discover because
(i) the cross section is enhanced by the resonance, and
(ii) there are only charged leptons without any missing energy in the final states, which give a clean signal even at the hadron collider.

Now our idea is to extend this dilepton $Z'$ resonance (Figure \ref{fig:diagrams} (a)) to a multi-lepton $Z'$ resonance and use it to search for other new physics.
If there is a process which has a new particle like Higgs or superpartner between the $Z'$ and charged leptons, it will be a good channel to search for the new particle at the LHC, exploiting the same advantages (i) and (ii).

The 4-lepton $Z'$ resonance with superparticles in the middle can serve as the SUSY search channel (Figure \ref{fig:diagrams} (b)).
The 6-lepton $Z'$ resonance with Higgs in the middle can serve as the Higgs search channel (Figure \ref{fig:diagrams} (c)).
We will check the feasibility of these multi-lepton $Z'$ resonances at the LHC with the design energy of 14 TeV.

\section{4-lepton $Z'$ resonance: SUSY search}
\begin{figure}[tb]
  \includegraphics[width=.33\textwidth]{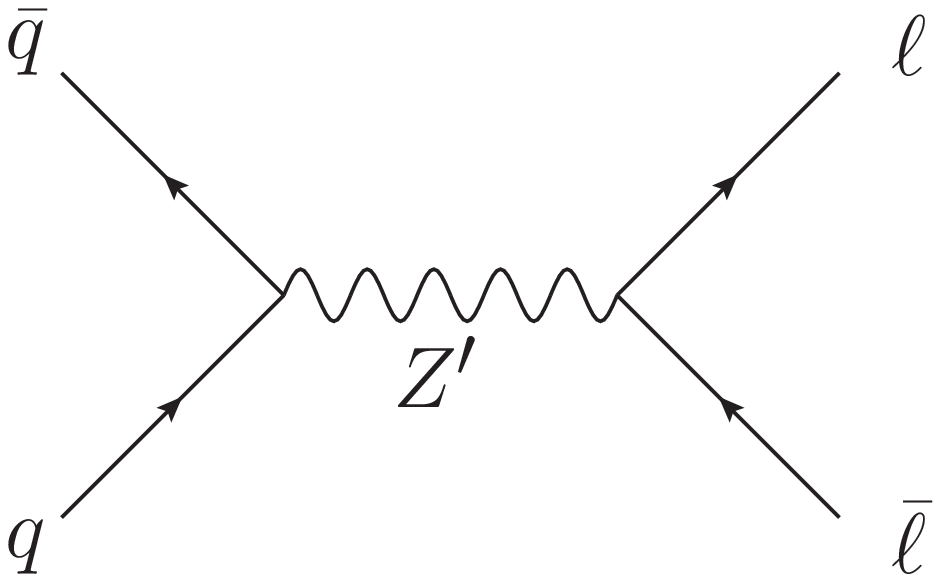}  ~
  \includegraphics[width=.33\textwidth]{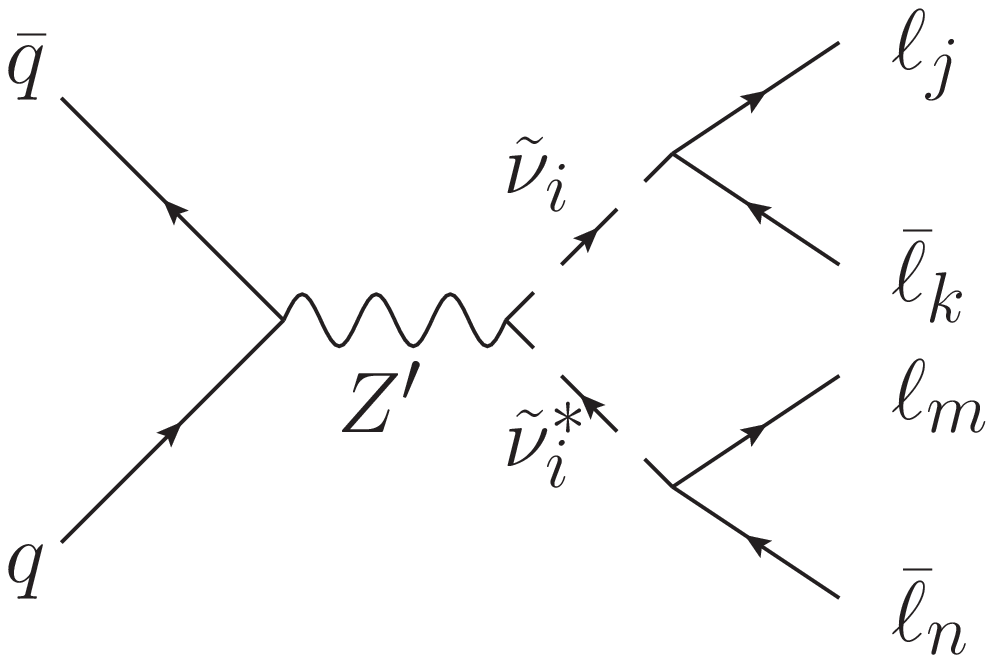}  ~~~
  \includegraphics[width=.33\textwidth]{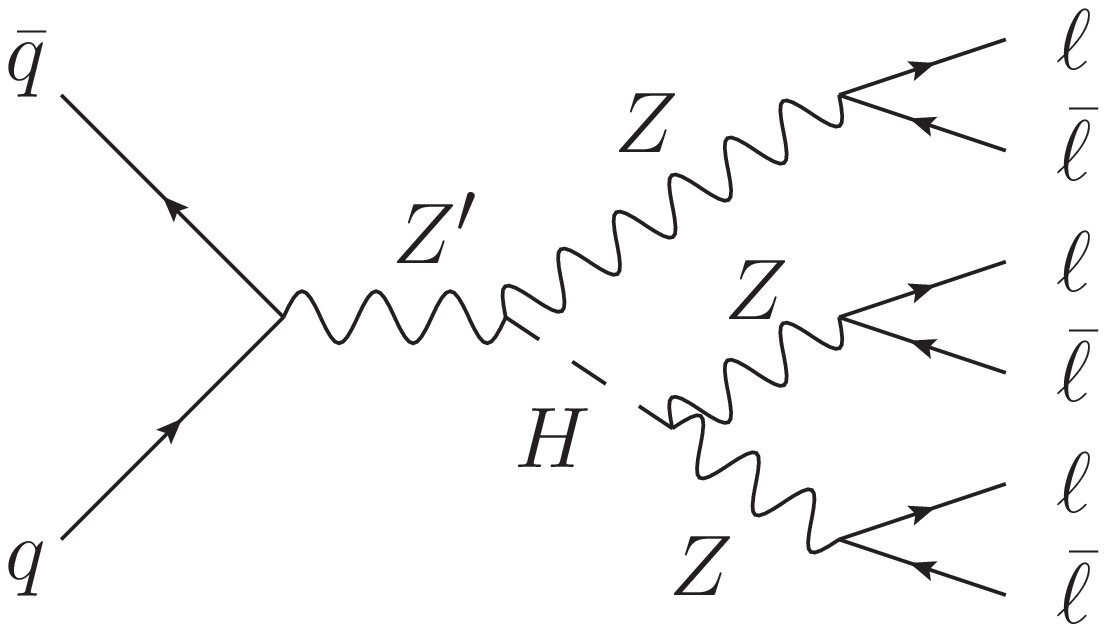}
  \caption{Various leptonic $Z'$ resonances as discovery channels: (a. left) $2\ell$ resonance ($Z'$ discovery), (b. center) $4\ell$ resonance (SUSY discovery), (c. right) $6\ell$ resonance (Higgs discovery).}
  \label{fig:diagrams}
\end{figure}

For details of this section including a choice of parameter values, see Ref.~\cite{Lee:2008cn}.

If the $Z'$ is discovered by the dilepton resonance, it will motivate the search for the $R$-parity violating signal since we may not need $R$-parity for the stability of the proton and dark matter in the SUSY framework any longer.
A TeV scale $U(1)'$ model, which has $B_3 \times U_2$ as its remnant discrete symmetry, can protect the proton and a hidden sector dark matter candidate although the lepton number ($\lnum$) is manifestly violated at tree level. (See Ref.~\cite{Lee:2008zzl} for a review of this scenario developed in Refs.~\cite{Lee:2007fw,Hur:2007ur,Lee:2007qx,Lee:2008pc,Hur:2008sy}.\footnote{This physics scenario was presented at SUSY 2008 conference in Seoul, Korea \cite{Lee:2008em}.}) 

We assume the LSP is a left-handed sneutrino ($\til\nu$).
We consider a 4-lepton $Z'$ resonance through the $\lnum$ violating term $\lambda LLE^c$ (Figure \ref{fig:diagrams} (b)).
\be
q \bar q \to Z' \to \til\nu \til\nu^* \to 4\ell ~~~~(\ell = e, \mu)
\ee

The $\lambda LLE^c$ term provides a relation between the $U(1)'$ charge of the left-handed leptons and right-handed leptons: $2 z[e_L] - z[e_R] = 0$.
This enforces, when $M_{Z'} \gg m_{\til\nu}$,
\be
\Gamma(Z' \to \til\nu \til\nu^*) \sim 0.1 \times \Gamma(Z' \to e^+ e^-) ,
\ee
which leads to
\be
\sigma_{4\ell} \simeq \sigma(p p \to Z') \br(Z' \to \til\nu \til\nu^*) \br(\til\nu \to 2\ell)^2 \sim [0.1 \times \sigma_{e^+ e^-}] \times \br(4\ell) , \label{eq:sigma4l}
\ee
where $\br(4\ell) \equiv \br(\til\nu \to 2\ell)^2$ is the branching fraction of the 4 light leptons from the $\til\nu$ LSP pair.

Though $\br(4\ell)$ depends on the texture of the $\lnum$ violating couplings and the $\til\nu$ LSP flavor, $\br(4\ell) \sim 0.1$ is a realistic value, which makes $\sigma_{4\ell} \sim 10^{-2} \times \sigma_{e^+ e^-}$ possible from Eq. \eqref{eq:sigma4l}.

\begin{figure}[tb]
  \includegraphics[width=.45\textwidth]{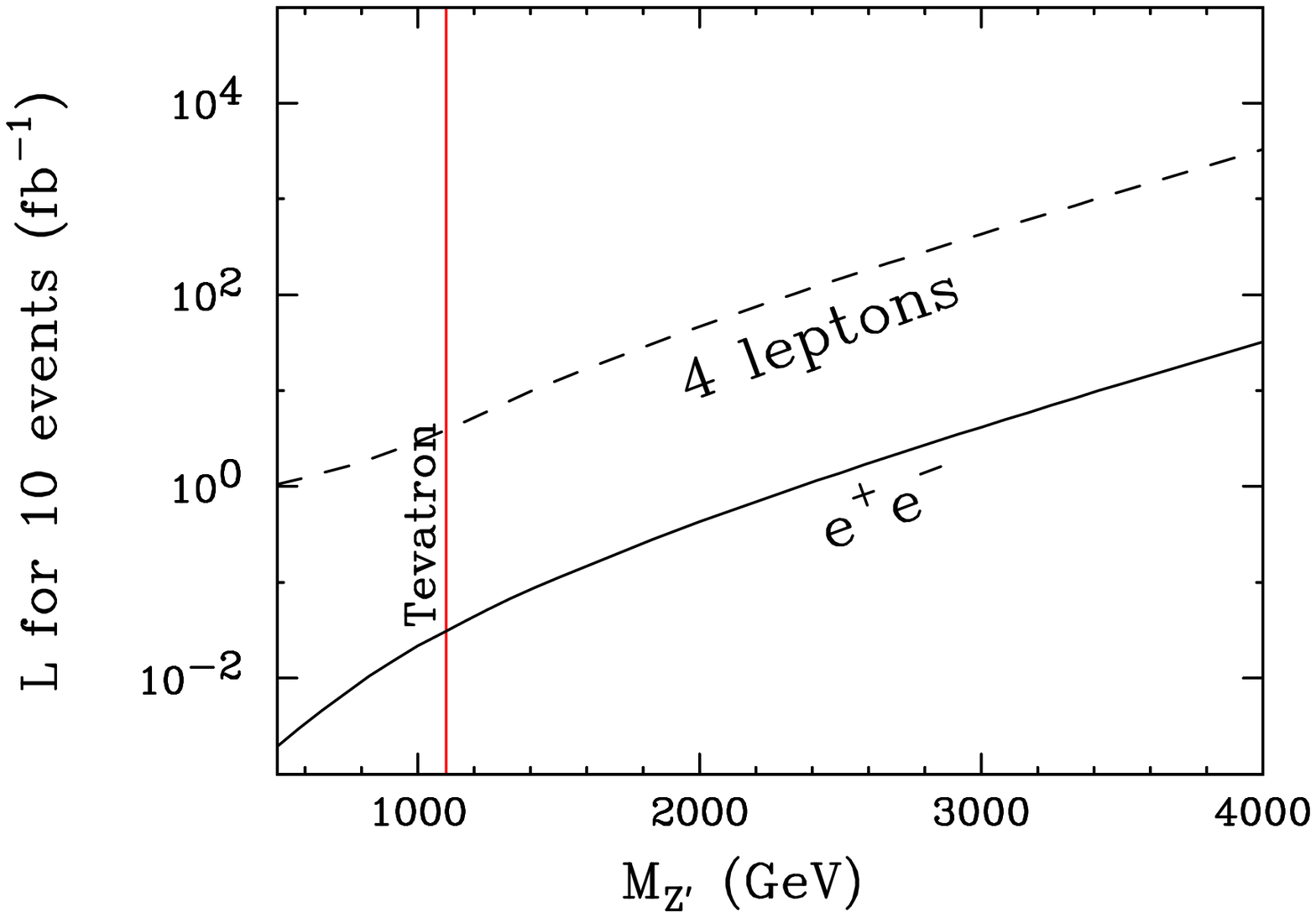} ~~~~~~~~
  \includegraphics[width=.45\textwidth]{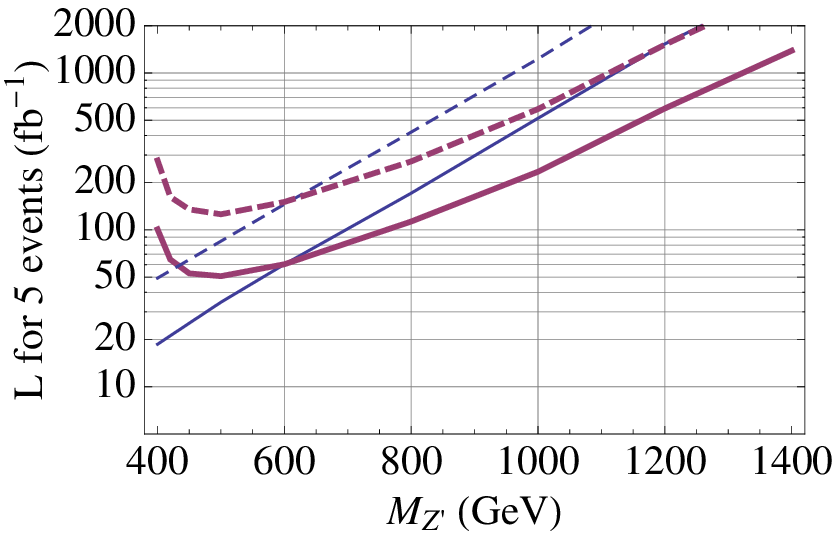}
  \caption{(a. left) Luminosity to discover $6$-lepton $Z'$ resonance. (b. right) Luminosity to discover 4-lepton $Z'$ resonance.}
  \label{fig:luminosity}
\end{figure}

Figure \ref{fig:luminosity} (a) shows the luminosity to get 10 events of dielectron resonance (solid) and 4-lepton resonance (dashed) for a certain $Z'$ coupling.
(Irreducible background for a leptonic resonance is small.)
Luminosity of ${\cal O}(1 \sim 10) ~\fb^{-1}$ may be sufficient to discover the 4-lepton resonance.

It should be noted that a $\ell$ should be charged under the $U(1)'$ if a $\til\nu$ is charged.
Thus a dilepton $Z'$ resonance is unavoidable when there is a 4-lepton $Z'$ resonance.
If a dilepton resonance is found at the LHC, we better look for the 4-lepton resonance with the same invariant mass.
If we are lucky, we may be able to see the SUSY signal in the early stage of the LHC.

\section{6-lepton $Z'$ resonance: Higgs search}
For details of this section including a choice of parameter values, see Ref.~\cite{Barger:2009xg}.

Now we want to move on to the 6-lepton $Z'$ resonance.
It does not require SUSY although $U(1)'$ and Higgs have a strong connection under SUSY \cite{Barger:2006dh}.

We consider a 6-lepton $Z'$ resonance through $Z'$-$Z$-$H$ coupling (Figure \ref{fig:diagrams} (c)).
\be
q \bar q \to Z' \to Z H \to Z Z Z \to 6\ell ~~~~(\ell = e, \mu)
\ee

The $Z' \to Z H$ channel has been studied in an $E_6$ context \cite{Barger:1987xw,Gunion:1987jd,Deshpande:1988py,Hewett:1988xc}.
$Z'$-$Z$-$H$ coupling depends on the details of the Higgs spectrum and their mixing angle, but just for the illustration, we want to look at the single Higgs doublet case.
\be
L_{\rm kin} = \left|\left(\partial_\mu - \frac{i}{2} g_Z Z_\mu + i g_{Z'} z[H] Z'_\mu \right) \frac{1}{\sqrt{2}} (H+v) \right|^2 = -g_Z g_{Z'} z[H] v H Z_\mu Z'^\mu + \cdots,
\ee
where $g_Z$, $g_{Z'}$, $z[H]$, and $v$ are the coupling constants of the $Z$, and $Z'$, the $U(1)'$ charge of the Higgs doublet, and the Higgs vacuum expectation value, respectively.
Therefore $Z'$-$Z$-$H$ coupling can be sizable if the $U(1)'$ charge of the Higgs doublet is sizable.

We assume $m_H > 2 M_Z$ so that all intermediate particles are on-shell.
\be
\sigma_{6\ell} \simeq \sigma(pp \to Z') \br(Z' \to Z H) \br(H \to Z Z) \br(Z \to \ell \bar\ell)^3
\ee
$\br(Z' \to Z H) \times \br(H \to Z Z)$ is model dependent, but ${\cal O}(1\%)$ is possible in our example.
The major suppression comes from triple leptonic $Z$ decay.

It should be noted that this process does not require a direct coupling of the $Z'$ to the leptons.
Therefore, it is possible to have this leptonic $Z'$ resonance even for a leptophobic $Z'$, serving as a discovery channel for such a $Z'$.
This suggests that the 6-lepton $Z'$ resonance search should be performed regardless of the dilepton search results, even for the $M_{Z'}$ disfavored by the dilepton search data.

Figure \ref{fig:luminosity} (b) shows the luminosity to get 5 events of the 6-lepton resonance.
Luminosity of ${\cal O}(10 \sim 100) ~\fb^{-1}$ may be enough for the discovery.

\section{Summary}
Higgs is probably the most important discovery goal at the LHC, and SUSY is arguably the next most important goal.
A TeV scale $U(1)'$ gauge symmetry is well motivated, for example, as an alternative to R-parity in the SUSY framework.

The $U(1)'$ has interesting implications including various multi-lepton $Z'$ resonances, which are clean signals at the LHC:
(i) the dilepton $Z'$ resonance is the best way to search for the $Z'$,
(ii) the 4-lepton $Z'$ resonance can help the SUSY search,
(iii) the 6-lepton $Z'$ resonance can help the Higgs search.

In short, a multi-lepton $Z'$ resonance is a great venue to search for important new physics.


\begin{theacknowledgments}
It is my pleasure to thank all my collaborators for enjoyable collaborations.
This work was supported by the Department of Energy under grant DE-FG03-94ER40837.
\end{theacknowledgments}



\bibliographystyle{aipproc}   

\bibliography{sample}

\begin{thebibliography}{9}

\bibitem{Lee:2008cn}
  H.~S.~Lee,
  Phys.\ Lett.\  B {\bf 674}, 87 (2009)
  [arXiv:0812.1854 [hep-ph]].

\bibitem{Barger:2009xg}
  V.~Barger, P.~Langacker and H.~S.~Lee,
  arXiv:0909.2641 [hep-ph].

\bibitem{Langacker:2008yv}
  P.~Langacker,
  arXiv:0801.1345 [hep-ph].

\bibitem{Langacker:2009im}
  P.~Langacker,
  arXiv:0909.3260 [hep-ph].

\bibitem{Lee:2008zzl}
  H.~S.~Lee,
  Mod.\ Phys.\ Lett.\  A {\bf 23}, 3271 (2008)
  [arXiv:0811.2539 [hep-ph]].

\bibitem{Lee:2007fw}
  H.~S.~Lee, K.~T.~Matchev and T.~T.~Wang,
  Phys.\ Rev.\  D {\bf 77}, 015016 (2008)
  [arXiv:0709.0763 [hep-ph]].

\bibitem{Hur:2007ur}
  T.~Hur, H.~S.~Lee and S.~Nasri,
  Phys.\ Rev.\  D {\bf 77}, 015008 (2008)
  [arXiv:0710.2653 [hep-ph]].
  
\bibitem{Lee:2007qx}
  H.~S.~Lee, C.~Luhn and K.~T.~Matchev,
  JHEP {\bf 0807}, 065 (2008)
  [arXiv:0712.3505 [hep-ph]].

\bibitem{Lee:2008pc}
  H.~S.~Lee,
  Phys.\ Lett.\  B {\bf 663}, 255 (2008)
  [arXiv:0802.0506 [hep-ph]].
  
\bibitem{Hur:2008sy}
  T.~Hur, H.~S.~Lee and C.~Luhn,
  JHEP {\bf 0901}, 081 (2009)
  [arXiv:0811.0812 [hep-ph]].

\bibitem{Lee:2008em}
  H.~S.~Lee,
  AIP Conf.\ Proc.\  {\bf 1078}, 569 (2009)
  [arXiv:0808.3600 [hep-ph]].

\bibitem{Barger:2006dh}
  V.~Barger, P.~Langacker, H.~S.~Lee and G.~Shaughnessy,
  Phys.\ Rev.\  D {\bf 73}, 115010 (2006)
  [arXiv:hep-ph/0603247].
  
\bibitem{Barger:1987xw}
  V.~D.~Barger and K.~Whisnant,
  Phys.\ Rev.\  D {\bf 36}, 3429 (1987).

\bibitem{Gunion:1987jd}
  J.~F.~Gunion, L.~Roszkowski and H.~E.~Haber,
  Phys.\ Rev.\  D {\bf 38}, 105 (1988).

\bibitem{Deshpande:1988py}
  N.~G.~Deshpande and J.~Trampetic,
  Phys.\ Lett.\  B {\bf 206}, 665 (1988).

\bibitem{Hewett:1988xc}
  J.~L.~Hewett and T.~G.~Rizzo,
  Phys.\ Rept.\  {\bf 183}, 193 (1989).

\end{thebibliography}

\IfFileExists{\jobname.bbl}{}
 {\typeout{}
  \typeout{******************************************}
  \typeout{** Please run "bibtex \jobname" to optain}
  \typeout{** the bibliography and then re-run LaTeX}
  \typeout{** twice to fix the references!}
  \typeout{******************************************}
  \typeout{}
 }



\end{document}